\documentclass[11pt]{article}

\pdfoutput=1

\usepackage{graphicx}
\usepackage{fullpage}
\usepackage{setspace}
\usepackage{amssymb}
\usepackage{amsmath}
\usepackage{bm}
\usepackage{subcaption}
\usepackage{url}
\usepackage[square, comma, numbers, sort&compress]{natbib}
\usepackage[left=1in,top=1in,right=1in,bottom=1in,nohead]{geometry}
\usepackage{listings}
\begin{document}

\title{Automatic Differentiation of a Finite-Volume-Based Transient Heat Conduction Code for Sensitivity Analysis}

\begin{center}
\thispagestyle{empty}
~
\vfill
{\Large \textbf{Automatic Differentiation of a Finite-Volume-Based Transient Heat Conduction Code for Sensitivity Analysis}}
\vfill
\textbf{Christopher T. DeGroot}$^{1*}$\\
cdegroo5@uwo.ca
\vfill
$^1$Department of Mechanical and Materials Engineering\\
Western University\\
London, ON, N6A 5B9, Canada
\vfill
\textbf{Running head:} Automatic Differentiation of Heat Conduction Code
\vfill
\end{center}


\newpage
\onehalfspacing

\begin{abstract}
A general method for computing derivatives of solution fields and other simulation outputs, with respect to arbitrary input quantities, is proposed. The method of automatic differentiation is used to carry out differentiation and propagate derivatives through the simulation code by chain rule, in forward order. An object-oriented approach using the operator overloading and templating features of the C++ programming language is presented. Verification results are given for a plane wall with surface convection, where the derivative of the dimensionless temperature field with respect to the Biot number is computed and compared to an analytical solution. Further results are given for conduction in a composite material with regions of different thermal conductivity. The derivative of the temperature field is computed with respect to the conductivity of one of the phases using the proposed method.
\end{abstract}

\newpage
\section*{Nomenclature}

\renewcommand\arraystretch{1.5}

\begin{table}[h]
\begin{tabular}{ll}
$A$ & area, [$m^2$] \\
$A_n$ & coefficient in series expansion \\
$Bi$ & Biot number \\
$C_n$ & coefficient in series expansion \\
$\mathbf{D_{P,ip}}$ & displacement vector from $P$ to $ip$, [$m$] \\
$\mathbf{D_{nb,ip}}$ & displacement vector from $nb$ to $ip$, [$m$] \\
$Fo$ & Fourier number \\
$h$ & convection coefficient, [$W/m^2K$] \\
$H$ & dimension of domain, [$m$] \\
$\mathbf{J}$ & Jacobian matrix \\
$k$ & thermal conductivity, [$W/mK$] \\
$L$ & length of domain, [$m$], or differential operator \\
$\mathbf{n}$ & unit-normal vector \\
$N$ & number of control volumes in domain \\
$N_{ip}$ & number integration points surrounding a control volume \\
$r$ & residual at a control volume \\
$R$ & radius, [$m$] \\
$\mathbf{r}$ & residual vector \\
$t$ & time, [$s$] \\
$\Delta t$ & timestep, [$s$] \\
$T$ & temperature, [$K$] \\
$V$ & volume, [$m^3$] \\
\end{tabular}
\end{table}%

\newpage
\subsection*{Greek Symbols}
\begin{table}[h]
\begin{tabular}{ll}
$\alpha$ & thermal diffusivity, [$m^2/s$]\\
$\Delta\bf{\phi}$ & solution correction\\
$\zeta_n$ & eigenvalue \\
$\theta$ & dimensionless temperature \\
$\bm{\phi}$ & discrete solution field\\
$\phi^*$ & continuous solution field\\
$\psi$ & input parameter \\
$\partial\Omega$ & denotes a surface bounding a space $\Omega$ \\
$\Omega$ & denotes a three-dimensional space
\end{tabular}
\end{table}%

\subsection*{Subscripts and Superscripts}
\begin{table}[h]
\begin{tabular}{ll}
$\prime$ & denotes differentiation with respect to parameter $\psi$\\
$f$ & fiber \\
$i$ & iteration number \\
$ip$ & quantity evaluated at or associated with an integration point \\
$m$ & matrix \\
$n$ & timestep number \\
$nb$ & quantity evaluated at or associated with a control volume neighboring $P$ \\
$P$ & quantity evaluated at or associated with a control volume centroid\\
$\infty$ & quantity associated with ambient condition \\
\end{tabular}
\end{table}%

\newpage
\section{Introduction}
\label{sec:introduction}

Computational fluid dynamics (CFD) analysis is now generally regarded as a mainstream design tool, finding applications in many areas of heat transfer, including nanofluid flows, thermal energy storage, and many others.  While conventional CFD has become a reasonably mature field, an emerging area of interest involves prediction of solution sensitivities in addition to the solution of the primary system \cite{Jemcov2004,Jemcov2009,Colin2006,Duvigneau2006,Bischof2007}. For the purposes of this work, the term `sensitivity' broadly encompasses any derivative of a calculation output with respect to a calculation input.  This definition includes, as in the case of geometric shape optimization, the derivative of an integrated output quantity of interest with respect to the boundary location. Sensitivity analysis is, however, much more broad than geometric shape optimization, and can be used for uncertainty analysis and other types of optimization problems.

Solution sensitivities can be computed using a large variety of methods, perhaps the most obvious being by finite differences. Using finite differences, each input is perturbed by a small amount, the solution is recalculated, and the derivative is estimated using a divided difference. This method, however, suffers from truncation error if the difference in inputs is too large or cancellation error if the difference between inputs is too small \cite{Bischof1996,Griewank2008}.  Although finite differences are known to be useful and practical in many situations, it is of interest to find more accurate and efficient methods. Derivatives could also be calculated by manually coding the derivative of each function within the code and propagating the derivatives through the code by chain rule. Although this method could provide machine-accurate derivatives and can be written to be very efficient, it requires significant programming effort to produce and maintain.  Further, manually coding derivatives does not generalize to computing sensitivities with respect to arbitrary input variables and is prone to coding error. The continuous sensitivity equation (CSE) approach involves formally deriving solution sensitivities by implicit differentiation of the respective governing equations and the solution of the additional equations \cite{Colin2006,Duvigneau2006}. Although this approach is quite elegant and generalizes to arbitrary sensitivity parameters, the solution of the sensitivity equations does need to be implemented manually in the code and would require additional maintenance effort as the primary code is updated. Symbolic differentiation is another option that is available, but is not practical for large-scale computations where the complexity of the symbolic expressions would be too great \cite{Griewank2008}. The response surface methodology is another way to explore the sensitivity of a solution to its inputs by conducting a series of numerical simulations \cite{Shirvan2016,Wu2017}.

Automatic differentiation, which will be explained further in the following paragraphs, has many of the advantages that hand-coded derivatives have; namely, machine-accuracy and efficiency that is comparable to the main code.  It also generalizes to arbitrary input variables, as in the CSE approach, with the added advantage of minimal additional programming effort to implement and maintain. Automatic differentiation shares some similarities with symbolic differentiation, namely, that both are based on systematic application of chain rule to all operations within the given code. The main difference is that automatic differentiation applies chain rule to numerical values, while symbolic differentiation applies it to expressions \cite{Griewank2008}. For CFD applications, symbolic derivatives are not required, therefore automatic differentiation becomes a natural choice.

Automatic differentiation can be implemented in a number of different ways, including source code transformation \cite{Bischof1996,Giering1998,Bischof2007,Hascoet2013} and operator overloading \cite{Jemcov2004,Jemcov2009}. Source code transformation methods involve parsing of the base code to produce a secondary code that computes the required derivatives. While this approach can work well with procedural languages such as Fortran and C, it does not lend itself well to object oriented languages, such as C++ and Java. This is mainly due to the more complex class structures that are difficult to parse for generation of the derivative code.  Automatic differentiation can be implemented in object-oriented programming languages through the use of specialized classes that are designed to compute derivatives using overloaded operators. Using this approach, all functions and classes must be templated to work with arbitrary data types so that they can be used with the type that implements automatic differentiation. Although this could be a significant development task for a mature CFD code, it is straightforward to implement if done from the start of development. Since object-oriented languages, such as OpenFOAM \cite{Weller1998,Jasak2007} and SU2 \cite{Palacios2013,Palacios2014,Economon2016}, are gaining popularity within the CFD community, this work focuses on the object-oriented approach for implementation of automatic differentiation, which could potentially be implemented in those codes.

In addition to the different implementation strategies discussed, automatic differentiation also has two main varieties, namely the forward mode and the reverse (adjoint) mode. The primary difference between the two modes is the direction in which the chain rule is applied. In forward mode, the relevant input quantities are flagged at the beginning of the simulation and the chain rule is propagated in forward order through the code. This results in the derivatives of all quantities with respect to the selected input(s). In reverse mode, a specific output quantity is identified and all operations and intermediate values within the code are recorded. Once the simulation is complete, the chain rule is propagated in backwards order, starting with the selected output quantity and results in the derivative of that quantity with respect to all input variables. The advantage of the reverse mode is that its cost is independent of the number of input parameters, making it ideal for shape optimization, since the boundary locations are all considered as inputs. The advantage of forward mode is that it provides the derivatives of all calculated quantities within the code, including the solution fields at all locations. This makes forward mode ideal for uncertainty quantification and optimizations involving a small number of unknowns. While there are a number of studies of adjoint methods in CFD, there are relatively few involving forward mode, particularly for heat transfer applications.

The goal of this work is to implement the automatic differentiation technique in forward mode to a finite-volume-based code for analysis of transient heat conduction problems. The object-oriented approach will be used to enable sensitivity derivatives of any arbitrary parameter to be computed. The remainder of this article will be outlined as follows. First, the numerical methods will be outlined. This begins with the general formulation for the solution to a general partial differential equation, its linearization, and the formulation of the solution for the sensitivity derivatives using automatic differentiation. Then, the specific discretization of the transient heat conduction equation will be outlined. Next, results will be presented for code verification. The verification problem involves transient conduction in a plane wall with convection at the surface, and will be compared to an analytical solution for the dimensionless temperature field and the sensitivity of the solution with respect to the Biot number. Results will also be presented for transient conduction in a two-dimensional fiber-reinforced composite material, with sensitivities evaluated with respect to the thermal conductivity of the fibers. Finally, a discussion about the applicability of this technique to more general CFD codes will be provided.

\section{Numerical Methods}
\label{sec:numerical}

In this study, a finite-volume-based code is differentiated using the method of automatic differentiation, resulting in two separate solution fields that are solved by the code. The first, called the ``primary'' solution, is the solution to the governing partial differential equation(s) (PDE), and is the conventional solution that is normally obtained by the code without applying differentiation. The second solution field is the first derivative of the primary solution field with respect to a designated (arbitrary) input quantity.

In this section, the general primary solution strategy will be described first, followed by the first derivative solution. Subsequently, the implementation of automatic differentiation to find the required derivatives will be discussed. Finally, the governing equations and their discretization will be given.

\subsection{Primary Solution Method}

First, let us consider a PDE represented by the general differential operator
\begin{equation}
	L(\phi^*) = 0
    \label{eq:diff_operator_1}
\end{equation}
where $\phi^*$ represents the continuous solution to the PDE. To solve Eq.\ \ref{eq:diff_operator_1} using the finite-volume method, the continuous solution, $\phi^*$, is first approximated by a discrete solution vector $\bm{\phi} \in \mathbb{R}^N$, where $N$ is the number of control volumes in the discretized domain. The PDE is then integrated over each control volume and each term is approximated using the discrete solution $\bm{\phi}$.  In general, the solution will not exactly satisfy the discretized equation; rather it will have a residual, $\mathbf{r} \in \mathbb{R}^N$. To solve the system of equations, Newton's method is used to linearize the problem. To do so, the residual vector is expanded in a Taylor series about the solution at a particular iteration, $i$, denoted $\bm{\phi}_i$.  The goal is to find the solution where $\mathbf{r} = 0$. This procedure results in
\begin{equation}
    \mathbf{r}(\bm{\phi}_i)
    + \left.\frac{\partial\mathbf{r}}{\partial\bm{\phi}}\right|_{\bm{\phi}_i}
    \left(\bm{\phi} - \bm{\phi}_i \right)
    = 0
    \label{eq:resid_1}
\end{equation}
The Jacobian of the residual vector is denoted as
\begin{equation}
    \mathbf{J}(\bm{\phi})
    = \frac{\partial\mathbf{r}}{\partial\bm{\phi}}
    \label{eq:jacobian_1}
\end{equation}
which defines a linear system that is to be solved for the solution correction $\Delta\bm{\phi}=\left(\bm{\phi} - \bm{\phi}_i \right)$, according to
\begin{equation}
    \mathbf{J}\left(\bm{\phi}_i\right)
    \Delta\bm{\phi}
    = - \mathbf{r}(\bm{\phi}_i)
    \label{eq:linear_system_1}
\end{equation}
The solution is then updated according to the fixed point iteration
\begin{equation}
    \bm{\phi} = \bm{\phi}_i + \Delta\bm{\phi}_i
    \label{eq:fixed_point_1}
\end{equation}

\subsection{First Derivative Solution Method}

The first derivative of the solution vector $\bm{\phi}$, with respect to an arbitrary quantity $\psi$, is obtained by differentiating Eqs.\ \ref{eq:linear_system_1} and \ref{eq:fixed_point_1} with respect to $\psi$. This results in the linear system
\begin{equation}
    \mathbf{J}(\bm{\phi}_i)
    \Delta\bm{\phi}^\prime
    = - \mathbf{r}^\prime(\bm{\phi}_i)
    - \mathbf{J}^\prime(\bm{\phi}_i)\Delta\bm{\phi}_i
    \label{eq:linear_system_2}
\end{equation}
and the fixed point iteration
\begin{equation}
    \bm{\phi}^\prime
    = \bm{\phi}_i^\prime
    + \Delta\bm{\phi}_i^\prime
    \label{eq:fixed_point_2}
\end{equation}
where the prime symbol represents differentiation with respect to $\psi$. The solution sensitives, $\bm{\phi}^\prime$, are then obtained by iteratively solving the linear system given by Eq.\ \ref{eq:linear_system_2} and updating the sensitivity solution using the fixed point iteration defined by Eq.\ \ref{eq:fixed_point_2}, in conjunction with the solution of the primary linear system and the associated iteration. Note that upon convergence of the primary system, the final term on the right side of Eq.\ \ref{eq:linear_system_2} vanishes and the fixed point iteration for the secondary system becomes independent of the primary iterations.

\subsection{Automatic Differentiation}

The concept underlying the forward mode of automatic differentiation is that the entire computational algorithm can be considered as a series of differentiable elementary operations to which chain rule may be applied \cite{Jemcov2004,Jemcov2009,Griewank2008}. Consider a function $f$ which is a composition of the functions $f_0, f_1, ..., f_n$, defined as
\begin{equation}
	f = f_n \circ f_{n-1} \circ ... \circ f_1 \circ f_0
\end{equation}
The derivative, obtained by applying chain rule is
\begin{equation}
	\frac{\partial f}{\partial f_0} =
    \frac{\partial f_n}{\partial f_{n-1}}
    \frac{\partial f_{n-1}}{\partial f_{n-2}} ...
    \frac{\partial f_2}{\partial f_1}
    \frac{\partial f_1}{\partial f_0}
    \label{eq:chain_rule}
\end{equation}
Provided each of the functions $f_0, f_1, ..., f_n$ are differentiable, the chain rule can be propagated from right to left in Eq.\ \ref{eq:chain_rule}. To implement in a numerical code, it is simply required that each of the elementary functions $f_0, f_1, ..., f_n$ is differentiable.

Since the code developed for this study is object-oriented C++, the approach for propagating the derivatives through the code is based on operator overloading. This involves implementing a specifically-designed data type that propagates derivatives through the code by overloading all of the standard mathematical operators. The overloaded operators compute not only the value of each expression, but also carry through the derivative by chain rule. Then, all classes and functions are templated so that they may be used for arbitrary data types, including the new type that implements automatic differentiation. Each of these concepts will be further described in the two following subsections.

An open source software project, entitled \texttt{WEdiff} (Western Engineering Automatic Differentiation Library) has been created  to implement the approach described in the C++ programming language \cite{WEdff}. In addition, \texttt{WEdiff} provides Python bindings so that it can be used with that programming language as well. This library is used to implement the automatic differentiation procedure within the finite-volume CFD code for this study.

\subsubsection{Operator Overloading}
\label{sec:operator_overloading}

Operator overloading refers to the customization of the standard
operators defined by a programming language for operands of user-defined types. This feature can be utilized to perform automatic differentiation through an suitably defined type. Consider the class \texttt{FwdDiff}, partially shown in Fig.\ \ref{fig:fwd_diff_class}, which is similar in concept to that of Jemcov \cite{Jemcov2009} and the implementation in \texttt{WEdiff}. As shown under the ``Constructors'' comment block, the \texttt{FwdDiff} class can be constructed from a standard \texttt{double}, another instance of a \texttt{FwdDiff}, and can be extended to any other constructors that are needed. All of the standard arithmetic operators must also be overloaded (only the operators \texttt{*=} and \texttt{+=} are shown in Fig. \ref{fig:fwd_diff_class}). Further, access methods must be provided, where the function \texttt{val()} returns the value of the \texttt{FwdDiff} object, while the function \texttt{dx()} returns its derivative with respect to the designated quantity. Each of these values is held as private member data within the class.

As an example, consider the implementation of the multiplication-assignment operator for the \texttt{FwdDiff} class. This operator takes the value of a variable, multiplies it by another number, and assigns the new value to the original variable. If the original variable $x$ is multiplied by $y$, we denote the result $z = xy$. This expression is used to calculate the value and derivative, which is then assigned back to $x$ in the code. The derivative is calculated as
\begin{equation}
	\delta z
    =
    \frac{\partial z}{\partial x}\delta x + \frac{\partial z}{\partial y}\delta y
    =
	y\delta x + x\delta y
\end{equation}
which can be seen in the implementation shown in Fig.\ \ref{fig:fwd_diff_mult_assign}. In this expression, $\delta x$ and $\delta y$ represent the derivatives of these intermediate variables up to the point where the expression is evaluated, accumulated by chain rule.

\subsubsection{Templating}
\label{sec:templating}

A template is an entity in the C++ programming language that defines a family of classes or functions that are specialized at compile-time for the specific type that is required. Templating is utilized as a means to write generic code that can be used with the \texttt{FwdDiff} class, as well as built-in types, such as \texttt{double}.

Figure \ref{fig:fwd_diff_example} illustrates a simple example where templating is used to calculate the derivative of a simple function using the \texttt{FwdDiff} class. In Fig.\ \ref{fig:fwd_diff_example}(a), a basic function is shown, which accepts a double-precision value $x$, performs some operations, and returns another double-precision value. Figure \ref{fig:fwd_diff_example}(b) shows the same function, except that it has been templated for an arbitrary type \texttt{T}. If we let \texttt{T=double} then we have the same function that is shown in Fig. \ref{fig:fwd_diff_example}(a). However, the templated function can also accept any other type that defines all of the required operators and functions (i.e. multiplication, addition, \texttt{sin}, and \texttt{pow}). Figure \ref{fig:fwd_diff_example}(c) shows a simple example using this function where \texttt{T=FwdDiff}. First, a variable $x$ is created and its value is set to $2$. To denote that the derivative is taken with respect to this value its derivative is initialized to $1$. Then the function is simply called and the template parameter is deduced based on the type of $x$. Finally, the derivative is reported. Running this code will print a derivative value of $-4.98595$, which is identical to the exact value, within the given number of decimal places.

In the finite-volume code developed for this study, all functions and classes were templated in a similar manner to the example shown, such that derivatives are propagated through the code, as in the example above.

\subsection{Governing Equations and Discretization}

\subsubsection{Governing Equations}

The PDE governing transient heat conduction for a material with constant thermophysical properties is
\begin{equation}
	\frac{\partial T}{\partial t} = \alpha\nabla^2T
    \label{eq:energy}
\end{equation}
where $T$ is temperature, $t$ is time, and $\alpha$ is the thermal diffusivity.

\subsubsection{General Discretization}

Equation \ref{eq:energy} is discretized using the finite-volume method by integrating the equation over a discrete polyhedral control volume in space, denoted $\Omega_P$. Let $\Omega_P$ have volume $V_P$ and be bounded by the control surface $\partial \Omega_P$ which is the union of the discrete control surfaces $\partial \Omega_{ip}$. Each control surface has area $A_{ip}$, where $ip\in\{1,2,...,N_{ip} \}$ and $N_{ip}$ is the number of discrete control surfaces. Carrying out this procedure on Eq.\ \ref{eq:energy} results in

\begin{equation}
	\frac{\partial T_{P}}{\partial t} V_{P}
	= \sum_{ip=1}^{N_{ip}} \alpha \left. \nabla T \right|_{ip} \cdot \mathbf{n}_{ip} A_{ip}
	\label{eq:disc_energy_1}
\end{equation}
where divergence theorem has been used to convert the volume integral of the term on the right into a surface integral. The discretization given in Eq.\ \ref{eq:disc_energy_1} can be shown to be second-order accurate in space and time, provided the points $P$ and $ip$ are located at the centroids of the control volume and control surfaces, respectively, and all further interpolations are made with second-order accuracy.

\subsubsection{Transient Term}

The transient term is discretized using a second-order backward Euler scheme, except at the first timestep, where a first-order backward Euler scheme must be used.  The second-order scheme, for a constant timestep $\Delta t$, is expressed as
\begin{equation}
	\frac{\partial T_P}{\partial t} V_{P} =
	\frac{\frac{3}{2}T_P^{n} - 2T_P^{n-1} + \frac{1}{2}T_P^{n-2}}{\Delta t} V_{P}
    \label{eq:time_resid}
\end{equation}
where $n$ denotes the timestep index, with $n$ being the current timestep.

\subsubsection{Diffusion Term}

\begin{figure}[ht]
	\centering
	\includegraphics[width=400pt]{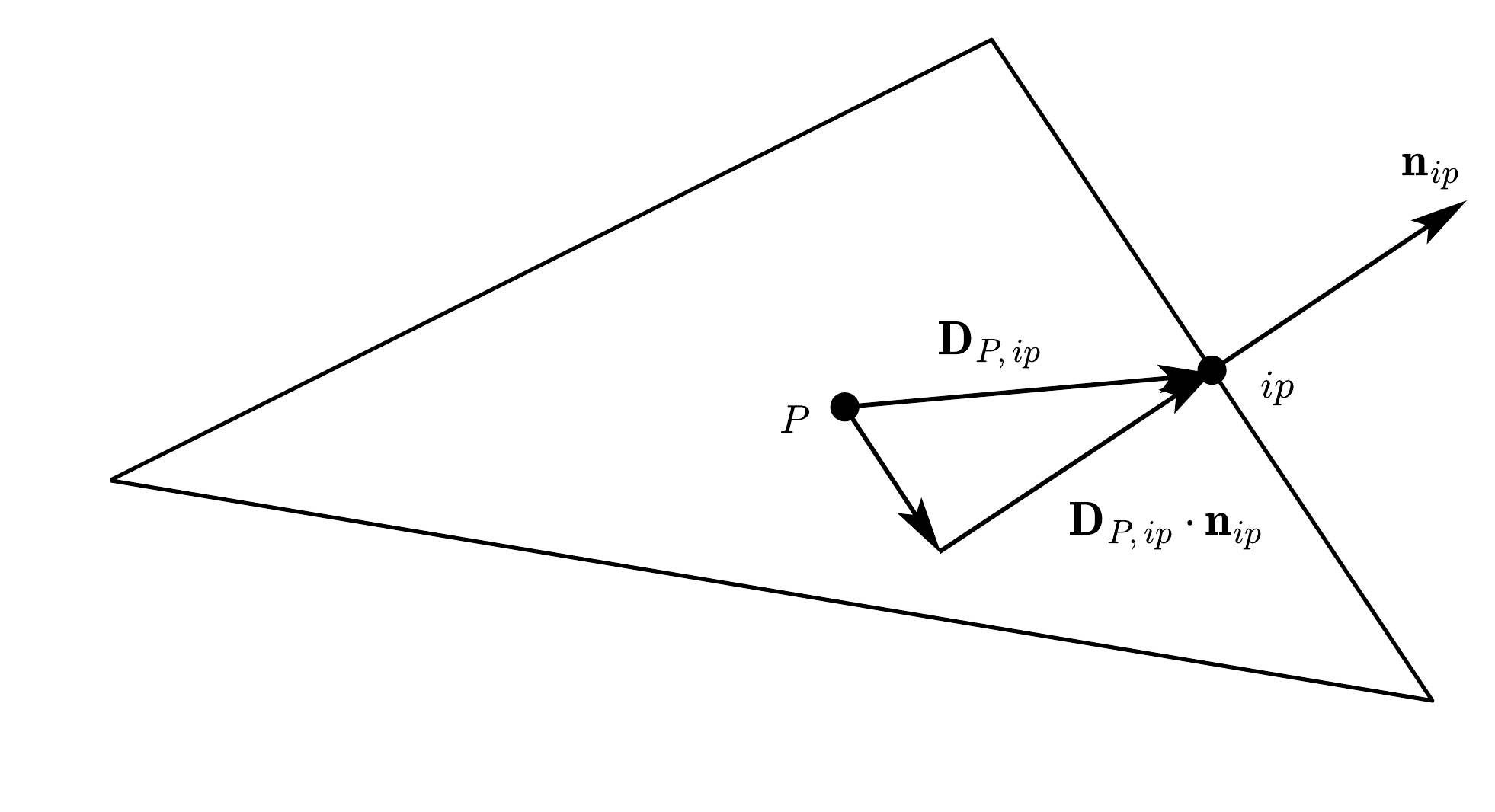}
	\caption{A control volume, $P$, with the relevant geometric parameters for evaluating a normal derivative noted.}
	\label{fig:cv_normal_derivative}
\end{figure}

Discretization of the diffusion term is based on the method presented by DeGroot and Straatman \cite{degroot2011}. This method was developed for fluid-porous interfaces, but is an effective general method to enforce a continuous flux across all shared faces within a finite-volume mesh, where thermal conductivity varies. The continuity of heat flux at a face is expressed mathematically as
\begin{equation}
	k_P \left. \nabla T \right|_{ip,P} \cdot \mathbf{n}_{ip}
    = k_{nb} \left. \nabla T \right|_{ip,nb} \cdot \mathbf{n}_{ip}
	\label{eq:diff_bal}
\end{equation}
where $nb$ refers to the neighboring control volume that shares the face containing the integration point $ip$. The derivatives normal to the integration point are computed by extrapolating from the cell-centers to a point located on a line that intersects $ip$ and is normal to the control surface, as shown in Fig. \ref{fig:cv_normal_derivative}. A finite-difference approximation can then be used to evaluate the normal derivative along this line as
\begin{equation}
	\left. \nabla T \right|_{ip,P} \cdot \mathbf{n}_{ip}
    = \frac{T_{ip} - \left[ T_P + \left. \nabla T \right|_P
	\cdot (\mathbf{D}_{P,ip} - (\mathbf{D}_{P,ip} \cdot \mathbf{n}_{ip})\mathbf{n}_{ip} ) \right]}{\mathbf{D}_{P,ip} \cdot \mathbf{n}_{ip}}
	\label{eq:norm_der_ip}
\end{equation}
Forming a similar expression for the control volume $nb$ and equating the two through Eq.\ \ref{eq:diff_bal}, results in the following expression for the integration point temperature that satisfies the heat flux from both sides of the control surface
\begin{multline}
	T_{ip} =
		\frac{k_{nb} (\mathbf{D}_{P,ip} \cdot \mathbf{n}_{ip})}
		{k_{nb} (\mathbf{D}_{P,ip} \cdot \mathbf{n}_{ip}) - k_P (\mathbf{D}_{nb,ip} \cdot \mathbf{n}_{ip})} T_{nb}
		- \frac{k_P (\mathbf{D}_{nb,ip} \cdot \mathbf{n}_{ip})}
		{k_{nb} (\mathbf{D}_{P,ip} \cdot \mathbf{n}_{ip}) - k_P (\mathbf{D}_{nb,ip} \cdot \mathbf{n}_{ip})} T_P \\
		+  \frac{k_{nb} (\mathbf{D}_{P,ip} \cdot \mathbf{n}_{ip})(\mathbf{D}_{nb,ip} - (\mathbf{D}_{nb,ip}\cdot\mathbf{n}_{ip})\mathbf{n}_{ip})}
		{k_{nb} (\mathbf{D}_{P,ip} \cdot \mathbf{n}_{ip}) - k_P (\mathbf{D}_{nb,ip} \cdot \mathbf{n}_{ip})} \cdot \left. \nabla T_P \right|_{nb} \\
		-  \frac{k_P (\mathbf{D}_{nb,ip} \cdot \mathbf{n}_{ip})(\mathbf{D}_{P,ip} - (\mathbf{D}_{P,ip}\cdot\mathbf{n}_{ip})\mathbf{n}_{ip})}
		{k_{nb} (\mathbf{D}_{P,ip} \cdot \mathbf{n}_{ip}) - k_P (\mathbf{D}_{nb,ip} \cdot \mathbf{n}_{ip})} \cdot \left. \nabla T_P \right|_P
	\label{eq:temperature_ip}
\end{multline}
Substituting Eq.\ \ref{eq:temperature_ip} back into Eq.\ \ref{eq:norm_der_ip} results in the following expression for the normal derivative, in terms of the cell-centered values, which ensures a flux balance across the control surface
\begin{multline}
	\left. \nabla T \right|_{ip,P} \cdot \mathbf{n}_{ip} =
	\frac{T_{nb} - T_P}{(\mathbf{D}_{P,ip}\cdot\mathbf{n}_{ip}) - \frac{k_P}{k_{nb}}(\mathbf{D}_{nb,ip}\cdot\mathbf{n}_{ip})}
	+ \frac{(\mathbf{D}_{nb,ip}-(\mathbf{D}_{nb,ip}\cdot\mathbf{n}_{ip})\mathbf{n}_{ip})}
	{(\mathbf{D}_{P,ip} \cdot \mathbf{n}_{ip}) - \frac{k_P}{k_{nb}}(\mathbf{D}_{nb,ip} \cdot \mathbf{n}_{ip})} \cdot\left.\nabla T \right|_{nb} \\
	- \frac{(\mathbf{D}_{P,ip}-(\mathbf{D}_{P,ip}\cdot\mathbf{n}_{ip})\mathbf{n}_{ip})}
	{(\mathbf{D}_{P,ip} \cdot \mathbf{n}_{ip}) - \frac{k_P}{k_{nb}}(\mathbf{D}_{nb,ip} \cdot \mathbf{n}_{ip})} \cdot\left.\nabla T \right|_P
	\label{eq:norm_der_ip_2}
\end{multline}
The two gradient terms in Eq.\ \ref{eq:norm_der_ip_2} account for non-orthogonality in the grid. In the limit of a completely orthogonal grid, Eq.\ \ref{eq:norm_der_ip_2} reduces to the harmonic mean formulation of Patankar \cite{Patankar1980}.

\subsubsection{Linearization}

The residual for a control volume $P$ is defined as
\begin{equation}
	r_P = \frac{\partial T_{P}}{\partial t} V_{P}
	- \sum_{ip=1}^{N_{ip}} \alpha \left. \nabla T \right|_{ip} \cdot \mathbf{n}_{ip} A_{ip}
	\label{eq:disc_residual}
\end{equation}
which vanishes at convergence. Each term in the residual equation is linearized in order to form the Jacobian matrix and assemble the global matrix system, as defined in Eq.\ \ref{eq:linear_system_1}. The secondary linear system, defined in Eq.\ \ref{eq:linear_system_2} is assembled by extracting the appropriate derivatives from values of type \texttt{FwdDiff}, which are computed automatically. All linear systems are solved using the linear solvers provided by the PETSc library \cite{petsc-efficient,petsc-web-page,petsc-user-ref}.

\section{Results and Discussion}
\label{sec:results}

\subsection{Plane Wall with Convection}

\subsubsection{Problem Description and Analytical Solution}

\begin{figure}[ht]
	\centering
	\includegraphics[width=400pt]{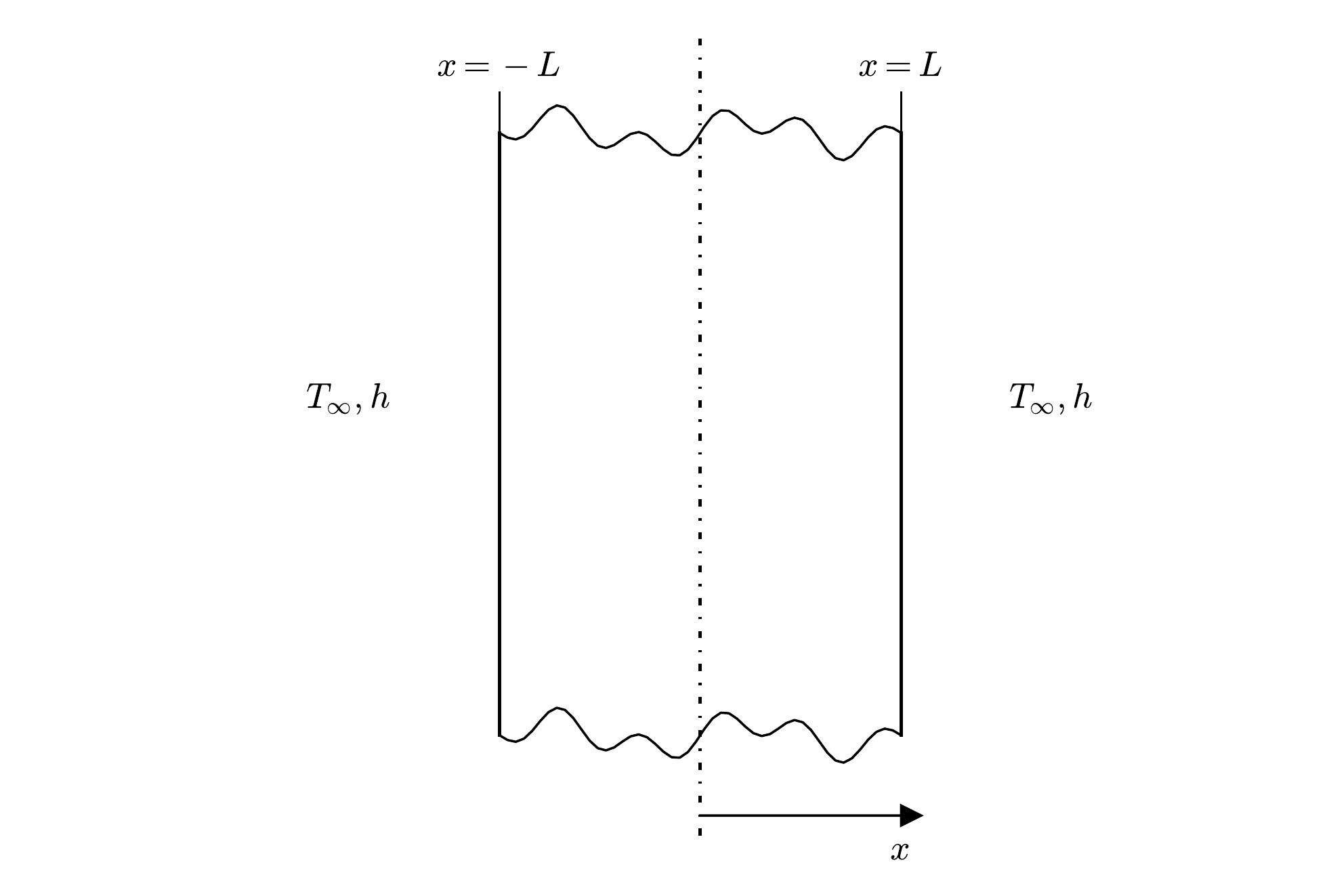}
	\caption{Schematic diagram of the plane wall with convection problem.}
	\label{fig:plane_wall_schematic}
\end{figure}

Consider a plane wall with thickness $2L$, as shown in Fig.\ \ref{fig:plane_wall_schematic}. The wall has an initially uniform temperature of $T_i$ when it is exposed to a fluid that results in a convective heat flux from the outer surfaces. The magnitude of the surface heat flux is $q_c=h\left[T(L,t) - T_\infty \right]$, where $h$ is a constant convection coefficient and $T_\infty$ is the ambient temperature of the fluid.

This problem has an analytical solution for the temperature field, $T$, as a function of time, which takes the form of the following infinite series \cite{Incropera2007}
\begin{equation}
	\theta =
    \frac{T-T_\infty}{T_i-T_\infty} =
    \sum_{n=1}^\infty
    C_n
    \exp\left(-\zeta_n^2 Fo\right)
    \cos\left(\zeta_n x^*\right)
    \label{eq:wall_analytical}
\end{equation}
where $x^*=x/L$ is the dimensionless distance from the centerline. Time is represented by the Fourier number, $Fo$, which is defined as
\begin{equation}
	Fo = \frac{\alpha t}{L^2}
    \label{eq:fourier}
\end{equation}
The values $\zeta_n$ are the eigenvalues of the problem and are the positive roots of the following equation
\begin{equation}
	\zeta_n \tan\zeta_n = Bi
	\label{eq:zeta_eqn}
\end{equation}
where $Bi$ is the Biot number, defined as
\begin{equation}
	Bi = \frac{hL}{k}
    \label{eq:biot}
\end{equation}
where $k$ is the thermal conductivity of the wall. The coefficient $C_n$ is defined in terms of $\zeta_n$ as
\begin{equation}
	C_n = \frac{4\sin\zeta_n}{2\zeta_n + \sin\left( 2 \zeta_n \right)}
    \label{eq:cn_eqn}
\end{equation}

The analytical solution above may also be differentiated with respect to a parameter of interest to determine the sensitivity of the solution to that parameter. In this case, the sensitivity of the dimensionless temperature field with respect to the Biot number, $Bi$, will be analyzed by differentiating the analytical solution, which can then be compared with the result from the differentiated numerical code.

To develop the analytical solution for the sensitivity with respect to $Bi$, let us first define
\begin{equation}
	A_n = C_n \exp\left(-\zeta_n^2 Fo\right)
\end{equation}
Then, differentiating Eq.\ \ref{eq:wall_analytical} with respect to $Bi$, we arrive at
\begin{equation}
	\frac{\partial \theta}{\partial Bi} =
    \sum_{n=1}^\infty
    \left[
    \frac{\partial A_n}{\partial Bi}
    \cos\left(\zeta_n x^*\right)
    - A_n
    \sin\left(\zeta_n x^*\right)
    \frac{\partial \zeta_n}{\partial Bi}
    x^*
    \right]
    \label{eq:deriv_T}
\end{equation}
where
\begin{equation}
	\frac{\partial A_n}{\partial Bi} =
    \left[
    \frac{\partial C_n}{\partial Bi}
    - 2 Fo C_n \zeta_n \frac{\partial \zeta_n}{\partial Bi}
    \right]
    \exp\left(-\zeta_n^2 Fo\right)
    \label{eq:deriv_An}
\end{equation}
\begin{equation}
	\frac{\partial C_n}{\partial Bi} =
    \left[
    \frac{4\cos\zeta_n}{2\zeta_n + \sin(2\zeta_n)}
    - \frac{4\sin\zeta_n\left[2\cos(2\zeta_n)+2\right]}{\left[2\zeta_n + \sin(2\zeta_n)\right]^2}
    \right]
    \frac{\partial \zeta_n}{\partial Bi}
    \label{eq:deriv_Cn}
\end{equation}
\begin{equation}
	\frac{\partial \zeta_n}{\partial Bi} =
    \frac{1}{\tan\zeta_n + \zeta_n \sec^2\zeta_n}
    \label{eq:deriv_zetan}
\end{equation}

\subsubsection{Numerical Solution}

The plane wall conduction problem is solved numerically using a grid of 50 uniform control volumes in the x-direction and a timestep of $10^{-3}$ [s]. The computational domain extends from $x=0$ to $x=L$ with a symmetry boundary condition applied at $x=0$ and a convective boundary condition at $x=L$.

\begin{figure}[ht]
	\centering
	\includegraphics[width=400pt]{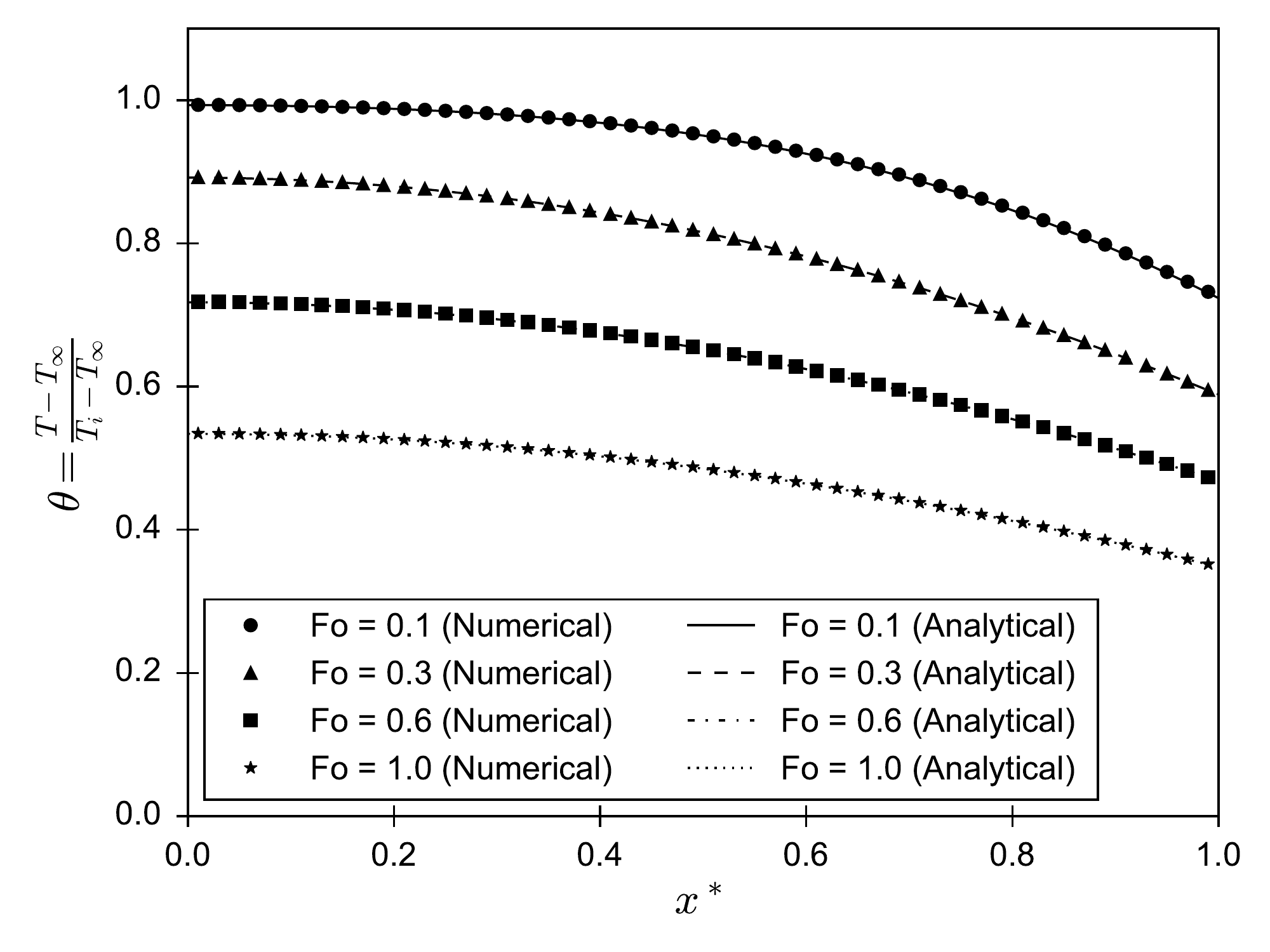}
	\caption{Comparison of the dimensionless temperature profiles for the plane wall with convection problem computed numerically and analytically.}
	\label{fig:plane_wall_temperature}
\end{figure}

The numerical solution for the dimensionless temperature field, $\theta$, is shown in Fig.\ \ref{fig:plane_wall_temperature} in comparison to the analytical solution given in Eq.\ \ref{eq:wall_analytical}. When computing the analytical solution, the first four terms in the series were used, although a single term is a good approximation for $Fo>0.2$ \cite{Incropera2007}. Results show that the numerical solution is nearly identical to the analytical solution, indicating that the code is functioning properly. As expected, the wall cools from the convective surface, which gradually cools the interior of the wall.

\begin{figure}[ht]
	\centering
	\includegraphics[width=400pt]{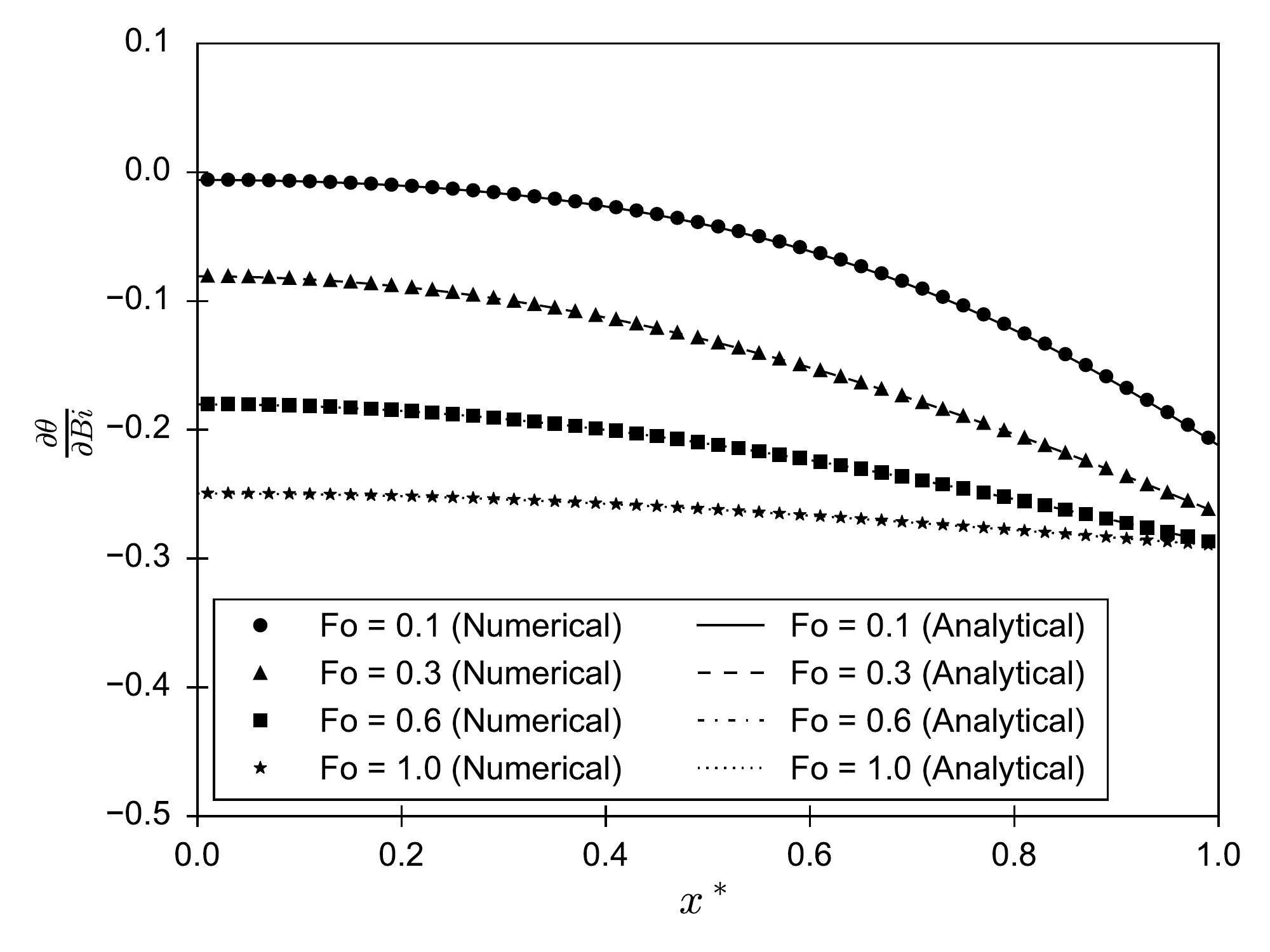}
	\caption{Comparison of the dimensionless profiles of the temperature derivative with respect to the Biot number, for the plane wall with convection problem, computed numerically and analytically.}
	\label{fig:plane_wall_temperature_derivative}
\end{figure}

The numerical solution for the derivative of $\theta$ with respect to $Bi$, obtained by automatic differentiation, is shown in Fig.\ \ref{fig:plane_wall_temperature_derivative} in comparison to the analytical solution given in Eqs.\ \ref{eq:deriv_T}-\ref{eq:deriv_zetan}. As for the dimensionless temperature field, the first four terms in the series were computed for the analytical solution. These results demonstrate that the sensitivity derivatives computed numerically are in excellent agreement with the analytical solution. Further analysis of Fig.\ \ref{fig:plane_wall_temperature_derivative} shows that the sensitivity to $Bi$ is always negative, due to the fact that increasing $Bi$ represents greater convective cooling, which leads to a lower temperature at a given time. As $Fo$ increases, the magnitude of the sensitivity increases, which is consistent with the fact that a change in $Bi$ has a greater cumulative effect as more time passes.

The results presented for the plane wall case have verified that the primary solution code as well as the sensitivity code are functioning properly. While this case is simple, it has demonstrated the fact that the derivative of a full solution field can be accurately calculated with respect to an arbitrary input value. It is worth noting that the sensitivity parameter could be taken to be any input (e.g.\ thermal conductivity, specific heat capacity, etc.) without any changes to the code. The sensitivity parameter is simply flagged when it is input into the code and all of its derivatives are propagated automatically to all other variables in the code.

\subsection{Conduction in a Fiber-Reinforced Composite}

\subsubsection{Problem Description}

\begin{figure}[ht]
	\centering
	\includegraphics[width=350pt]{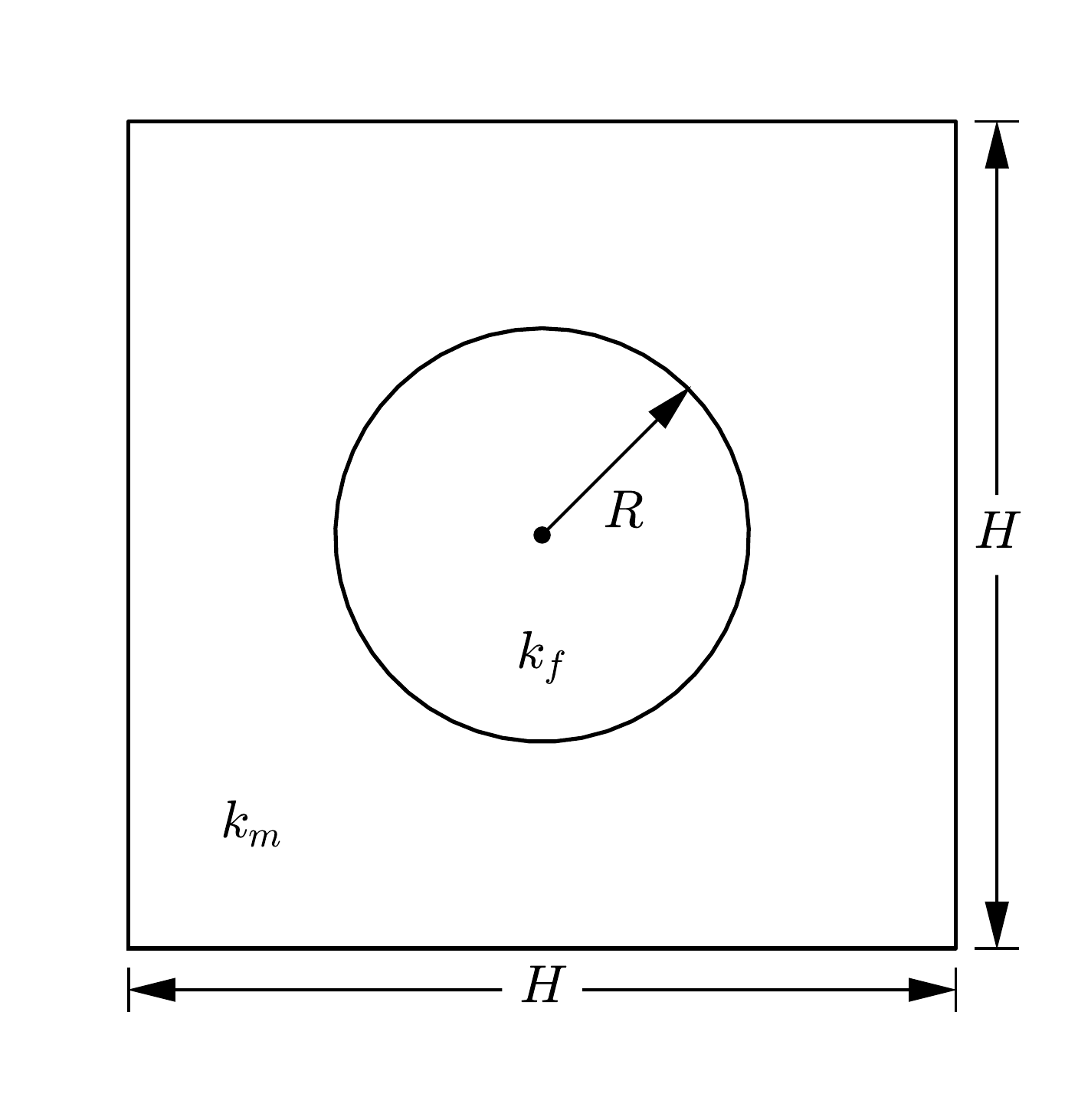}
	\caption{Schematic diagram of the fiber-reinforced composite problem.}
	\label{fig:composite_schematic}
\end{figure}

Conduction in a simple fiber-reinforced composite material is considered, where the fiber has a thermal conductivity $k_f$ and the matrix has a thermal conductivity $k_m$. The fiber has a radius $R$ and is contained with a square section of the matrix with dimension $H$, as shown in Fig.\ \ref{fig:composite_schematic}. Similar fibers are assumed to be arranged periodically in the horizontal direction, while the upper and lower surfaces are held at different constant temperatures, $T_U$ and $T_L$, respectively. The fibers are assumed to be very long such that the problem is effectively two-dimensional. For the problem under consideration $H=2$ mm and $R = 0.5$ mm.

\subsubsection{Numerical Solution}

\begin{figure}[ht]
	\centering
	\includegraphics[width=350pt]{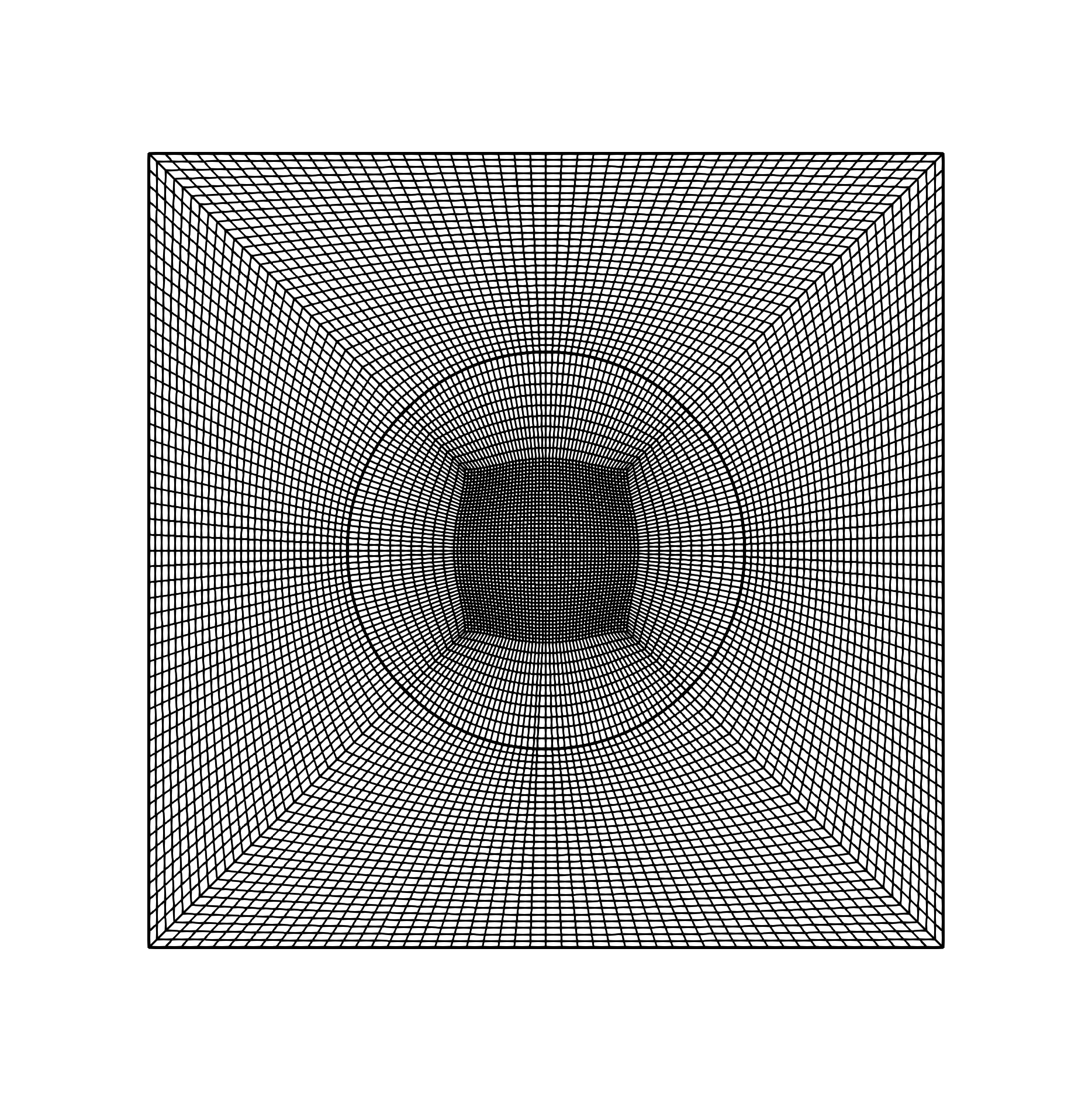}
	\caption{Plot of the computational mesh used to compute solutions for the fiber-reinforced composite problem.}
	\label{fig:composite_mesh}
\end{figure}

The problem is solved to steady state using the computational mesh shown in Fig.\ \ref{fig:composite_mesh}, which has just a single control volume into the page. On the left, right, front, and back surfaces, symmetry boundary conditions are applied. On the lower and upper walls, the temperatures $T_U$ and $T_L$ are specified as boundary conditions. The sensitivity of the temperature field is evaluated with respect to the fiber thermal conductivity, $k_f$.

\begin{figure}[ht]
  \begin{subfigure}[b]{0.5\linewidth}
  	\centering
	\includegraphics[width=\linewidth]{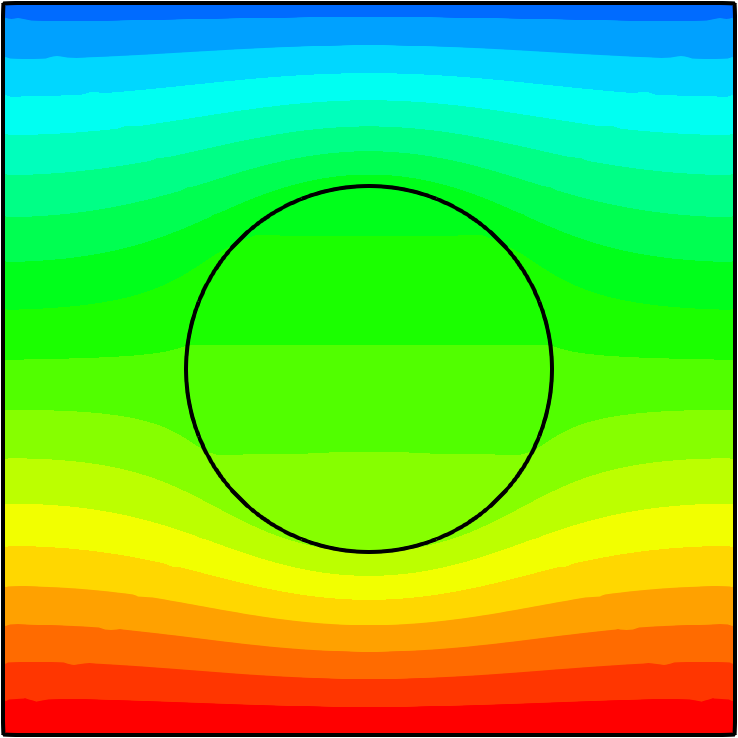}
    \includegraphics[height=35pt]{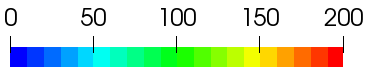}
    \caption{}
  \end{subfigure}
  \begin{subfigure}[b]{0.5\linewidth}
  	\centering
	\includegraphics[width=\linewidth]{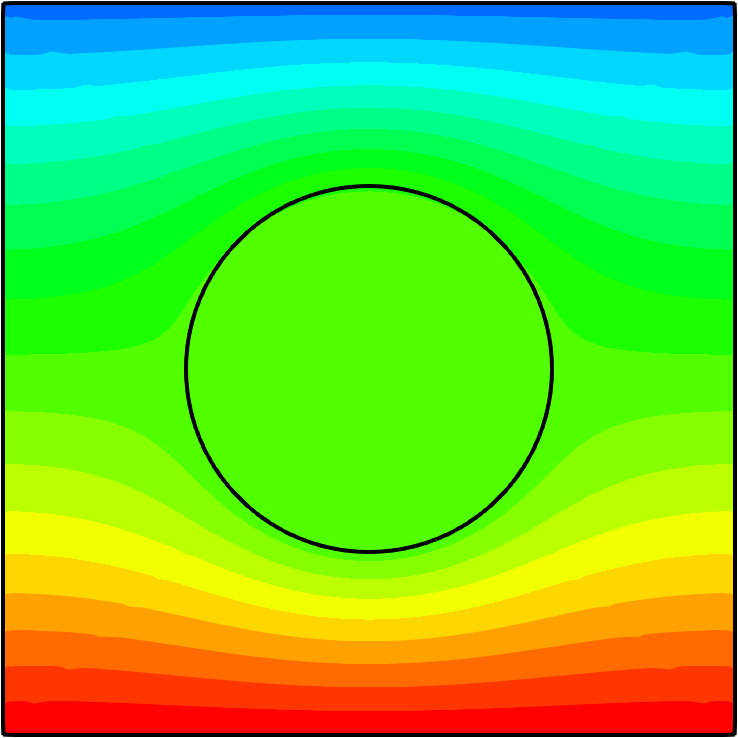}
    \includegraphics[height=35pt]{TempLegend}
    \caption{}
  \end{subfigure}
  \vspace{12pt}
  \begin{subfigure}[b]{0.5\linewidth}
  	\centering
	\includegraphics[width=\linewidth]{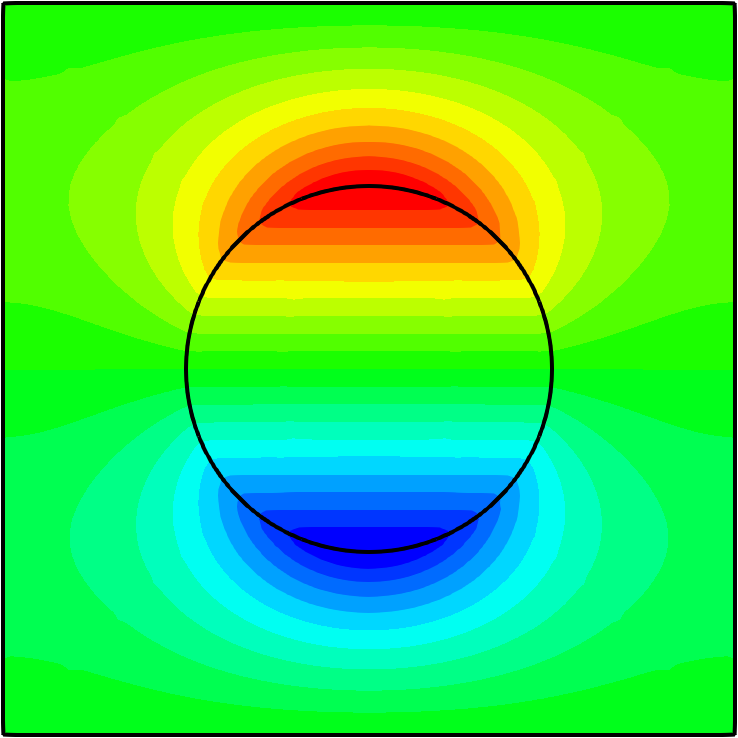}
    \includegraphics[height=35pt]{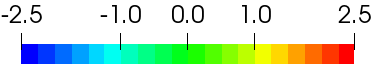}
    \caption{}
  \end{subfigure}
  \begin{subfigure}[b]{0.5\linewidth}
  	\centering
	\includegraphics[width=\linewidth]{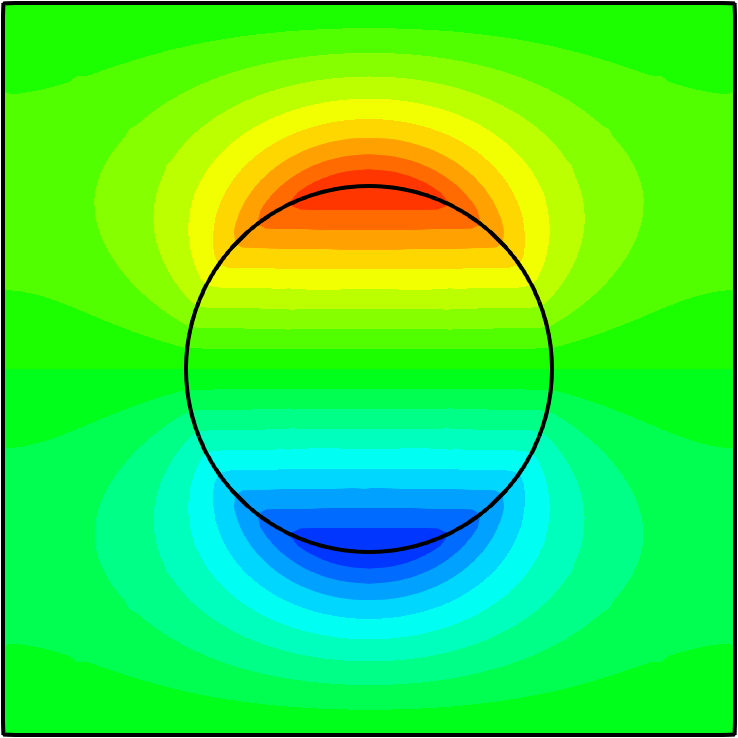}
    \includegraphics[height=35pt]{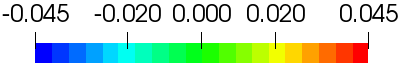}
    \caption{}
  \end{subfigure}
\caption{Contour plots of the temperature fields [(a) and (b)] and derivatives of temperature with respect to fiber thermal conductivity [(c) and (d)] for the cases of $k_f = 5 \text{ W/m}\cdot\text{K}$ [(a) and (c)] and $k_f = 50 \text{ W/m}\cdot\text{K}$ [(b) and (d)].}
\label{fig:composite_results}
\end{figure}

Results are shown in Fig.\ \ref{fig:composite_results} for cases where $T_U = 25^\circ\text{C}$, $T_L = 200^\circ\text{C}$, and $k_m= 1 \text{ W/m}\cdot\text{K}$. Two different values for the thermal conductivity of the fiber are tested, namely $k_f \in \{5, 50\} \text{ W/m}\cdot\text{K}$. Examination of the temperature fields indicates that changing the fiber conductivity has only a minor effect on the overall temperature field. When the conductivity is higher, the fiber is closer to becoming isothermal. When it is lower, it can be seen that a small temperature gradient develops within the fiber. A major difference can be seen in the derivative fields, noting that the two contours are plotted using different scales because of their vastly different magnitudes. In both cases, the shape is similar, indicating that increasing fiber conductivity leads to lower temperatures on the hot side and higher temperatures on the cold side of the fiber. The magnitudes of the derivatives show, however, that there is minimal effect once the conductivity has reached $50 \text{ W/m}\cdot\text{K}$. When the conductivity is just $5 \text{ W/m}\cdot\text{K}$, compared with the matrix conductivity of $1 \text{ W/m}\cdot\text{K}$, there is a larger effect of increasing fiber conductivity.

This case demonstrates in greater detail the types of results that can be obtained by running heat transfer simulations that include an automatic differentiation feature. Simulations of this type could be used to estimate the effects of changing material properties, optimize material selection, or be used to assess uncertainties in simulation results based on uncertainties in thermophysical properties. Again, the choice of the sensitivity parameter is arbitrary and can be changed without changing any of the underlying code, making this method very generally applicable.

\section{Conclusions}
\label{sec:conclusions}

This study has presented a general method for computing sensitivity derivatives of simulation outputs with respect to arbitrary input quantities using forward mode automatic differentiation. The automatic differentiation technique was implemented using an object oriented approach, utilizing operator overloading and templating facilities within the C++ programming language. The method has been validated using analytical solutions for transient conduction in a plane wall with convection at the surface, demonstrating the accuracy of the method. Further results were presented for conduction in a fiber-reinforced composite material to further illustrate the capabilities of the method. While the present work focused on transient conduction, it is straightforward to extend the method into more general CFD calculations involving fluid flow, turbulence, and other transport processes, since there is minimal additional code required to implement the sensitivity solution, once the primary solution is implemented.

\section*{Acknowledgements} \nonumber
The author acknowledges the financial support provided by the Natural Sciences and Engineering Research Council (NSERC) under grant number RGPIN-2017-04078.

\newpage
\doublespacing
\bibliographystyle{nht}
\bibliography{references.bib}

\newpage
\onehalfspacing
\listoffigures

\newpage
\begin{figure}[ht]
\begin{lstlisting}[language=C++,frame=single]
class FwdDiff
{
public:
  // Constructors
  FwdDiff(const double x);
  FwdDiff(const FwdDiff& x);
  ...
  // Overloaded operators
  FwdDiff& operator*=(const FwdDiff& x);
  FwdDiff& operator+=(const FwdDiff& x);
  ...
  // Access
  double val() const;
  double dx() const;
  ...
private:
  // Value and derivative
  double _val;
  double _dx;
};
\end{lstlisting}
\caption{Partial definition of the interface for a class \texttt{FwdDiff} that is used to propagate derivatives using forward mode automatic differentiation.}
\label{fig:fwd_diff_class}
\end{figure}

\newpage
\begin{figure}[ht]
\begin{lstlisting}[language=C++,frame=single]
FwdDiff& operator*=(const FwdDiff& x)
{
  _dx = _dx*x.val() + _val*x.dx();
  _val *= x.val();
  return *this;
}
\end{lstlisting}
\caption{Implementation of the multiplication-assignment operator for the \texttt{FwdDiff} class.}
\label{fig:fwd_diff_mult_assign}
\end{figure}

\newpage
\begin{figure}[ht]
  \begin{subfigure}[b]{\linewidth}
    \begin{lstlisting}[language=C++,frame=single]
double f(const double& x)
{
  return x*sin(pow(x, 2.0)) + x;
}
    \end{lstlisting}
    \caption{}
    \vspace{12pt}
  \end{subfigure}
  \begin{subfigure}[b]{\linewidth}
    \begin{lstlisting}[language=C++,frame=single]
template <typename T>
T f(const T& x)
{
  return x*sin(pow(x, 2.0)) + x;
}
    \end{lstlisting}
    \caption{}
    \vspace{12pt}
  \end{subfigure}
  \begin{subfigure}[b]{\linewidth}
    \begin{lstlisting}[language=C++,frame=single]
// Create a variable of type FwdDiff
FwdDiff x;

// Initialize the independent variable
x.setVal(2.0);
x.setDx(1.0);

// Evaluate function
FwdDiff y = f(x);

// Report the derivative
std::cout << ``df/dx at x = 2.0: '' << y.dx() << std::endl;
    \end{lstlisting}
    \caption{}
  \end{subfigure}
\caption{An example of a templated function and the calculation of a derivative using the \texttt{FwdDiff} class for automatic differentiation.}
\label{fig:fwd_diff_example}
\end{figure}

\newpage

\newpage

\newpage

\newpage

\newpage

\newpage

\newpage

\end{document}